\documentclass[doublecol]{epl2}

\usepackage{graphicx}
\usepackage{dcolumn}
\usepackage{bm}
\usepackage{color}
\usepackage{float}
\usepackage{sistyle}
\usepackage{psfrag}

\newcommand{\drawat}[3]{\makebox[0pt][l]{\raisebox{#2}{\hspace*{#1}#3}}}

\renewcommand{\Im}{\mathrm{Im}}

\title{Dissipation of micro-cantilevers as a function of air pressure and metallic coating}
\shorttitle{Viscous and viscoelastic dissipation of AFM cantilevers} 

\author{T. J. LI\inst{1,2} \and L. BELLON\inst{1}\thanks{Email: \email{Ludovic.Bellon@ens-lyon.fr}}}
\shortauthor{T. J. Li \etal}

\institute{
  \inst{1} Universit\'e de Lyon, Laboratoire de physique, ENS Lyon, CNRS - 46 all\'e d'Italie, Lyon 69364, France\\
  \inst{2} Department of Physics, East China Normal University - 3663 Zhongshan North Rd., Shanghai 200062, China
}
 
\abstract{In this letter, we characterize the internal dissipation of coated micro-cantilevers through their mechanical thermal noise. Using a home-made interferometric setup, we achieve a resolution down to $\SI{E-14}{m/\sqrt{Hz}}$ in the measurement of their deflection. With the use of the fluctuation dissipation theorem and of the Kramers-Kronig relations, we rebuilt the full mechanical response function from the measured noise spectrum, and investigate frequency dependent dissipation as a function of the air pressure and of the nature of the metallic coatings. Using different thicknesses of gold coatings, we discuss the source of the internal viscoelastic damping.}

\begin{document}

\maketitle

Microsized cantilevers are present in many applications, ranging from chemical and biological sensors~\cite{Lavrik2004} to atomic force microscopy (AFM)~\cite{Meyer2004}. They are also used as a basic brick of microelectromechanical systems (MEMS). As mass detectors for example, they can be functionalized for the adsorption of specific chemical compounds by proper coating of their surface: the mass increment is detected as a resonance frequency shift. Gold coating is widely used due to the large selection of materials that can be adsorbed via thiol chemistry~\cite{Lavrik2004,Ziegler2004}.
Besides, from a gold layer, electrically conducting layers can be patterned and integrated into the cantilever for local heating~\cite{Ataka1993}, magnetomotive actuation~\cite{Lange2002} or piezoresistive readout of the deflection~\cite{Rasmussen2003}.

The sensitivity of the resonant cantilever when used as a mass sensor depends on the spectral resolution, thus of its quality factor $Q$ defined as the ratio of stored vibrational energy over energy lost per cycle of vibration~\cite{Yasumra2000}. The greatest sensitivity will thus be reached with the smallest dissipation. The functionality of MEMS, AFM probe or mass sensors is based on the deformation of the cantilever. Thermally induced mechanical fluctuations determine the ultimate deflection sensitivity of these sensors and represent one of the most important noise sources. They are linked to the damping of the system, as shown by the fluctuation dissipation theorem (FDT)~\cite{Callen1952}. Smaller damping will thus lead to reduced thermal noise, thus increasing the sensitivity and usability of the probes. It is therefore of prime importance to understand and characterize the dissipation sources in these systems.

Great effort has been put into describing the effect of ambient pressure and cantilever coating on their resonant behavior, both experimentally~\cite{Bergaud1999,Chon2000,Yum2004,Sandberg2005-A,Sandberg2005-B,Paolino2009} and theoretically~\cite{Saulson1990,Leveque1997,Sader1998}. For example, Sandberg~\cite{Sandberg2005-B} investigated the effect of gold coating on the quality factor of a resonant cantilever and showed that in vacuum $Q$ is severely reduced by the deposition of even a thin gold film (100nm), especially for higher order modes. Considering only structural damping, Saulson~\cite{Saulson1990} proposed a viscoelastic model, in which the power spectrum density (PSD) of thermal induced deflection presents a characteristic $1/f$ like trend. In a previous work~\cite{Paolino2009}, we introduced a simple power law to describe the frequency dependence of this viscoelasticity on a gold coated cantilever, and a model that includes Sader's approach to describe the coupling with the surrounding atmosphere~\cite{Sader1998,Bellon2008}. We showed that the damping is only due to the coating when viscous dissipation vanishes in vacuum. Understanding the source of this coating induced viscoelasticity will undoubtedly help in designing more sensitive and accurate cantilever based sensors and microprobes. However, measuring the thermal noise or small damping over a wide range of frequency is a great challenge, and very few experiments~\cite{Numata2003,Gonzlez1995,Kajima1999,Bellon2008,Paolino2009} have succeeded so far in directly measuring fluctuations out of resonance, notably at low frequency.

\begin{figure}[tbH]
\begin{center}
\includegraphics{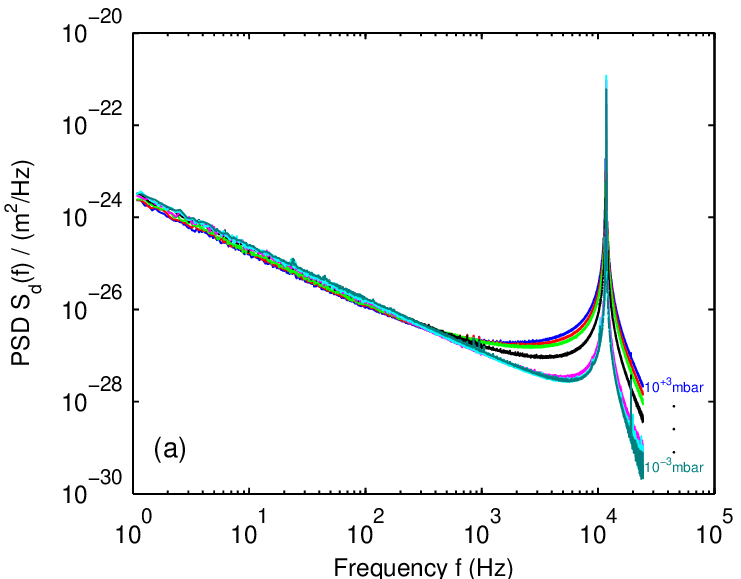}
\drawat{-49mm}{32mm}{
\psfrag{d}[Bl][Bl]{\small $d$}%
\psfrag{Er}[Bl][Bl]{}
\psfrag{Es}[Bl][Bl]{}
\includegraphics[width=0.15\textwidth]{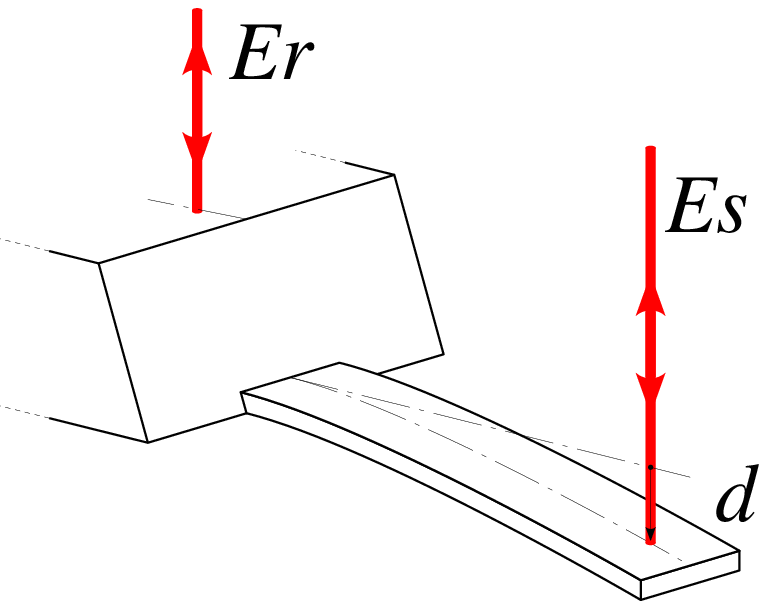}}
\includegraphics{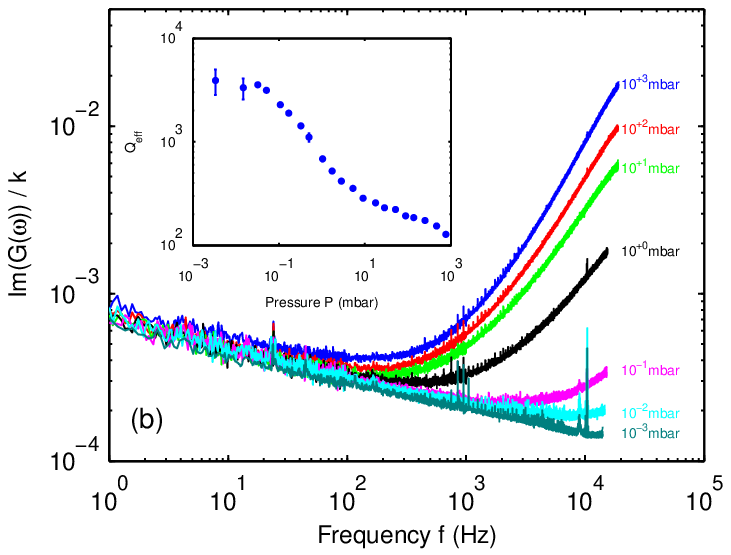}%
\end{center}
\caption{(a) Thermal noise spectrum and (b) reconstructed dissipative part of the response function $G$ of cantilever $\bf{cAu}$ at different pressures. Background noise due to the electronics has been subtracted from the measured PSD. Viscous damping, dominant at high frequency and high pressure over viscoelastic dissipation, vanishes in vacuum, leading to a sharper resonance. The inset in (a) illustrates our high precision interferometric method: interferences between beams reflecting on the base and free extremity of the cantilever directly measure its deflection $d$~\cite{Paolino2009,Schonenberger1989}. Inset in (b): effective quality factor $Q_{\mathrm{eff}}$ at resonance as a function of pressure. \label{Fig:AuPressure}}
\end{figure}

In this work, we measure the thermal noise of AFM cantilevers, with a highly sensitive interferometric technique~\cite{Bellon2008,Paolino2009}, allowing access to the whole spectrum from 1 Hz to 20 kHz with a background noise as low as $\SI{E-28}{m^2/Hz}$ (see fig.~\ref{Fig:AuPressure}). With the use of the FDT and of the Kramers-Kronig relations, we rebuild the full mechanical response function from the measured PSD~\cite{Paolino2009}. We can therefore characterize dissipation of micro-cantilevers as a function of the air pressure and of the metallic coatings. Let us first recall briefly the theoretical background for our analysis of thermal noise, before describing the experiments and concluding with a discussion on the results.

Since our cantilevers are in equilibrium at temperature $T$, the thermal fluctuations of their deflection $d$ are described by the FDT, relating the PSD to the mechanical response function $G$ of the system:
\begin{equation}
S_d(f)=-\frac{4k_BT}{\omega}\Im\left[\frac{1}{G(\omega)}\right] \label{eq:FDT}
\end{equation}
where $k_B$ is the Boltzmann constant, $\omega=2\pi f$ the pulsation corresponding to frequency $f$, and $\Im$ stands for the imaginary part of its argument. $G(\omega)$ is defined in the Fourier space as $G(\omega)=F(\omega)/d(\omega)$, where $F$ is the force coupled to $d$ in the Hamiltonian of the system. Considering both viscoelastic and viscous damping~\cite{Saulson1990,Paolino2009}, we have:
\begin{equation}
G(\omega)=k\left[1-\frac{\omega^2}{\omega_0^2}+i\left(\frac{\omega}{\omega_0 Q_{a}}+\phi\right)\right]
\label{equation5}
\end{equation}
with  $k^\ast=k(1+i\phi)$ the complex spring constant, $\omega_0$ the resonant pulsation and $Q_{a}$ the quality factor linked to viscous damping in air. At resonance, we observe an effective quality factor  $Q_{\mathrm{eff}}$ defined as:
\begin{equation}
\frac{1}{Q_{\mathrm{eff}}}=\frac{1}{Q_{a}}+\phi
\label{equationQeff}
\end{equation}
In the experiments, it is hard to characterize dissipation outside resonance: the imaginary part of $G$ is much smaller than its real part, and the forcing method has to be perfectly controlled to perform such measurement. We therefore use the FDT (eq.~\ref{eq:FDT}) to infer $\Im[1/G(\omega)]$ from the measurement of the thermal noise $S_{d}$, then Kramers-Kronig integral relations to rebuilt the full mechanical response function $G$ from the knowledge of this imaginary part. Using this method, we have shown that both $Q_{a}$ and $\phi$ are in fact weakly frequency dependent for the model to match the experimental observations~\cite{Paolino2009}. $Q_{a}(\omega)$ is described by Sader's approach for viscous damping~\cite{Sader1998}, and the viscoelasticity is following a simple power law with a small exponent: $\phi(\omega)\propto\omega^{\alpha}$, with $\alpha \approx -0.11$ for a commercial gold coated cantilever~\cite{Paolino2009}. In this letter, we demonstrate how viscous damping vanishes when the pressure decreases, and characterize the behavior of $\phi$ for different coatings.

We use silicon AFM cantilevers (Budget Sensors CONT-x) with the following geometry: length $450\pm10\ \mu m$, width $50\pm5\ \mu m$, thickness $2\pm1\ \mu m$. Their resonant frequency is around $\SI{13}{kHz}$. We present here data corresponding to three samples : the coating of cantilever $\bf{cAu}$ is $\SI{5}{nm}$ of titanium plus $\SI{70}{nm}$ of gold (both sides), that of cantilever $\bf{cAl}$ is $\SI{70}{nm}$ of aluminum, and that of cantilever $\bf{cPt}$ is $\SI{70}{nm}$ of PtIr5. In addition, we performed successive gold layers deposition on three other probes: cantilever $\bf{dAu}$ is initially equivalent to $\bf{cAu}$, cantilever $\bf{dPt}$ is initially equivalent to $\bf{cPt}$, and cantilever $\bf{dSi}$ is initially a raw silicon cantilever, prepared with a $\SI{2}{nm}$ Cr adhesion layer. Layer thickness increments were around $\SI{10}{nm}$ for the first 3 evaporations, then around $\SI{20}{nm}$.

We first study the influence of ambient pressure $P$ on cantilever $\bf{cAu}$. In fig.~\ref{Fig:AuPressure}, we plot the PSD of the deflection for different pressures from ambient to $\SI{E-3}{mbar}$, and the corresponding reconstructed response function (imaginary part only, normalized by spring constant $k$). Dissipation is clearly the sum of two contribution: pressure independent viscoelasticity $\phi(\omega)$, always dominant at low frequency, and viscous damping term $\omega/\omega_{0}Q_{a}(\omega)$, dominant at high frequency and high pressure, but vanishing in vacuum. At resonance, the effective quality factor thus increases when the pressure drops, but saturate to a finite value due to the internal dissipation. For this cantilever and the frequency range probed here, viscous damping becomes negligible when $P<\SI{E-2}{mbar}$. In vacuum, the dissipative part of the response function directly leads to the viscoelaticity: $\phi(\omega)=\Im[G(\omega)]/k$ when $Q_{a}\rightarrow\infty$.
A simple power law matches well the observed frequency dependence~\cite{Paolino2009}, with an exponent $\alpha_{Au} \approx -0.15$.

\begin{figure}[tbH]
\begin{center}
\includegraphics{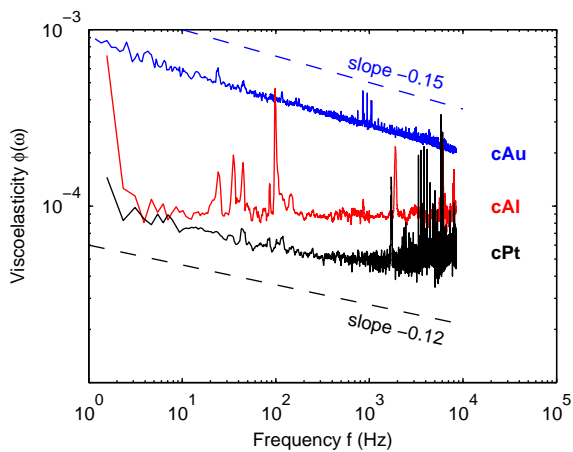}
\end{center}
\caption{Viscoelasticity $\phi(\omega)$ of cantilevers {\color{blue}$\bf cAu$}, {\color{red}$\bf cAl$} and {\color{black}$\bf cPt$} reconstructed from the noise spectrums measured in vacuum.\label{Fig:metals}}
\end{figure}

To check the generic validity of this behavior, we measure in vacuum the viscoelastic damping of different metallic coatings, and report the results for cantilevers $\bf{cAu}$, $\bf{cPt}$ and $\bf{cAl}$ on fig.~\ref{Fig:metals}. The magnitude of this internal dissipation is smaller for the 2 other coatings, but the weak frequency dependence is a common characteristic. At high frequency (above $\SI{1}{kHz}$), the viscous damping is still observable for cantilever $\bf{cPt}$ at the lowest pressure achievable in our system (\SI{E-3}{mbar}). If we limit the frequency range to $\SI{1}{kHz}$ for this sample, we observe a similar power law dependance with an exponent $\alpha_{Pt} \approx -0.12$. The data of the last cantilever, $\bf{cAl}$, is a a little bit different with a very flat measurement, leading to $\alpha_{Al} \approx 0$. The power law dependence thus works generically, though the exponent changes slightly for each cantilever.

\begin{figure}[tbH]
\begin{center}
\includegraphics{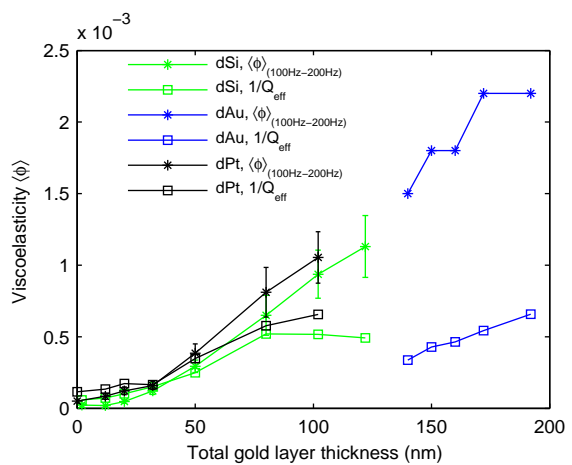}
\end{center}
\caption{Viscoelasticity at low and high frequency (average of $\phi(\omega)$ around $\SI{150}{Hz}$ and $1/Q_\mathrm{eff}$ at resonance) as a function of the total gold layer thickness for cantilevers $\bf{dSi}$, $\bf{dAu}$ and $\bf{dPt}$. Typical error bars (dispersion over a few acquisitions) are shown for two cantilevers.\label{Fig:thickness}}
\end{figure}

To investigate the origin of this internal damping, we finally measured the viscoelasticity of cantilevers $\bf{dSi}$, $\bf{dPt}$ and $\bf{dAu}$ as a function of the thickness of an added gold coating : the same 3 cantilevers where characterized in vacuum between successive layers deposition, and we plot in fig.~\ref{Fig:thickness} the value of the dissipation at low frequency (average value of $\phi(\omega)$ between $\SI{100}{Hz}$ and $\SI{200}{Hz}$) and at resonance (value of $1/Q_\mathrm{eff}$, deduced from a lorenzian fit of the PSD). Viscoelasticity remains weakly frequency dependent in these observations, and is roughly proportional to the thickness. This behavior suggests that the main contribution to the internal damping of the cantilever originates in the bulk of the coating, rather than from a surface or interface effect. Indeed, if such phenomenon was dominant, it would be present even for the lowest thickness, whereas we observe a vanishing dissipation when the coating layer gets thinner. Moreover, the value of the dissipation is independent of the adhesion layer for the gold coating : we get the same results for $\SI{2}{nm}$ of Cr (cantilever $\bf{dSi}$) and for $\SI{70}{nm}$ of PtIr5 (cantilever $\bf{dPt}$).

As a conclusion, let us summarize the main points illustrated in this letter. We measure the thermal noise of AFM cantilevers with a high resolution interferometer, and reconstruct from this data their dissipation. The damping of coated cantilevers is shown to arise from two distinct sources: viscous damping due to the external atmosphere, dominant at high frequency and ambient pressure, and internal viscoelastic damping, dominant at low frequency or in vacuum. This viscoelasticity is linked to the presence of a coating on the cantilever, and can be modeled by a complex spring constant. This dissipation is accurately fit by a simple power law $\omega^{\alpha}$ on up to 4 decades in frequency, with a small exponent $\alpha$ depending on coating material and procedure: for the cantilevers probed here, $\alpha_{Au} \approx -0.15$, $\alpha_{Pt}\approx -0.12$, and $\alpha_{Al}\approx 0$. Damping increases by one order of magnitude between PtIr5 and Au coating, Al falling in between those 2 materials. Finally, by changing the coating thickness, we show that dissipation is due to the bulk of the coating rather than to some interface friction between layers.

These results demonstrate that the choice of the coating material is critical with respect to internal dissipation in micro-cantilevers, and that gold is actually a bad choice with respect to this criterium. If this material is however required for its chemical properties, using the lowest functional thickness is advisable to minimize damping.

Our characterization procedure features an excellent resolution with measurement of mechanical loss tangents down to $\SI{e-4}{}$. This way, the viscoelasticity due to the coating can be accurately quantified and our measurements should be useful in the perspective of testing models of internal friction, eventually leading to improved coating procedures and better performance of cantilever based sensors.
Our method would also be suited to study other type of coatings, such as those implied in chemical or biological sensors, alone or linked to the target molecules.

\acknowledgments
We thank F. Vittoz and F. Ropars for technical support, and S. Ciliberto, A. Petrosyan and C. Fretigny for stimulating discussions.

\bibliographystyle{eplbib}
\bibliography{epl}

\end{document}